%% file: main.tex
\documentclass[lettersize,journal]{IEEEtran}

\usepackage[turnon]{notes}

\input{includes/includes}

\input{includes/pygments}

\input{includes/macros}

\title{LLM-based Test-driven Interactive Code Generation: User Study and Empirical Evaluation}

\begin{document}

\author{\IEEEauthorblockN{Sarah Fakhoury\IEEEauthorrefmark{1},
Aaditya Naik\IEEEauthorrefmark{2},
Georgios Sakkas\IEEEauthorrefmark{3},
Saikat Chakraborty\IEEEauthorrefmark{1} and
Shuvendu K. Lahiri\IEEEauthorrefmark{1} 
} \\
\IEEEauthorblockA{\IEEEauthorrefmark{1}Microsoft Research \\
\{sfakhoury, saikatc, shuvendu\}@microsoft.com\\
\IEEEauthorrefmark{2}University of Pennsylvania \\
asnaik@seas.upenn.edu \\
\IEEEauthorrefmark{2}University of California, San Diego \\
gsakkas@eng.ucsd.edu\\
}
}

\maketitle
\balance
\begin{abstract}
    \input{body/abstract}
\end{abstract}
\markboth{
}{Title}


\renewcommand\hl[1]{#1}
\input{body/introduction}

\input{body/related}

\input{body/method}

\input{body/userstudy}

\input{body/results}

\input{body/benchmark}

\input{body/threats}

\input{body/conclusion}




\bibliographystyle{IEEEtran}
\bibliography{main}

\end{document}

%% file: includes/includes.tex
\usepackage{hyperref}
\usepackage{boldline}
\usepackage{color}
\usepackage{float}
\usepackage{soul}
\usepackage[table]{colortbl}
\usepackage{tabularx}
\usepackage{bigstrut}
\usepackage[ruled]{algorithm2e}
\usepackage{amsmath,amsfonts}
\usepackage{amsmath}
\usepackage{booktabs}
\usepackage{graphicx} 
\usepackage{multirow}
\usepackage{array}
\usepackage{algorithmic}
\usepackage{balance}
\usepackage{blindtext}
\usepackage{caption}
\usepackage{hyperref}
\usepackage[noabbrev]{cleveref}
\usepackage{comment}
\usepackage{enumitem}
\usepackage{fancybox}
\usepackage{fancyvrb}
\usepackage{framed}
\usepackage{graphicx}
\usepackage{ifthen}
\usepackage{listings}
\usepackage{mathrsfs}
\usepackage{mdwmath}
\usepackage{mdwtab}
\usepackage{multirow}
\usepackage{pifont}
\usepackage{textcomp}
\usepackage{url}
\usepackage{xspace}
\usepackage[normalem]{ulem}
\usepackage[framemethod=TikZ]{mdframed}
\usepackage{caption}
\usepackage{subcaption}
\usepackage{tcolorbox}
\tcbuselibrary{listings,skins}
\usepackage{mathtools}
\usepackage{blindtext}
\usepackage{lstautogobble}
\DeclareGraphicsExtensions{.pdf,.jpeg,.png}
\graphicspath{{figures/}}
\colorlet{lightgrey}{lightgray}

\newlength\Linewidth
\def\findlength{\setlength\Linewidth\linewidth
\addtolength\Linewidth{-4\fboxrule}
\addtolength\Linewidth{-3\fboxsep}
}
\newenvironment{examplebox}{\par\begingroup
  \setlength{\fboxsep}{3pt}\findlength
  \setbox0=\vbox\bgroup\noindent
  \hsize=0.90\linewidth
  \begin{minipage}{0.90\linewidth}\normalsize}
    {\end{minipage}\egroup
    \textcolor{gray20}{\fboxsep1.2pt\fbox
     {\fboxsep5pt\colorbox{gray05}{\normalcolor\box0}}}
    \endgroup\par\noindent
    \normalcolor\ignorespacesafterend}

\usetikzlibrary{shadows}
\usepackage{graphics}
\newmdenv[
    tikzsetting= {fill=blueish},
    skipabove=0.33em,
    skipbelow=0.33em,
    linewidth=1pt,
    innerleftmargin=4pt,
    innerrightmargin=4pt,
    innertopmargin=2pt,
    innerbottommargin=2pt,
    linecolor=gray95,
    roundcorner=2pt, 
    shadow=true,
    shadowsize=4pt,
    shadowcolor=gray95
]{questionbox}

\newmdenv[
    tikzsetting= {fill=greenish},
    skipabove=0.33em,
    skipbelow=0.33em,
    linewidth=1pt,
    innerleftmargin=4pt,
    innerrightmargin=4pt,
    innertopmargin=2pt,
    innerbottommargin=2pt,
    linecolor=gray95,
    roundcorner=2pt, 
    shadow=true,
    shadowsize=4pt,
    shadowcolor=gray95
]{answerbox}

\newmdenv[
    skipabove=0.33em,
    skipbelow=0.33em,
    innerleftmargin=4pt,
    innerrightmargin=4pt,
    innertopmargin=2pt,
    innerbottommargin=2pt,
]{lessonbox}

\usepackage{tikz}
\newenvironment{lesson}
{
    \begin{lessonbox}
}
{\end{lessonbox}}

\newenvironment{result}
{\begin{answerbox}}
{\end{answerbox}}

\newenvironment{question}
{\begin{questionbox}}
{\end{questionbox}}

\definecolor{javared}{rgb}{0.6,0,0} 
\definecolor{javagreen}{rgb}{0.25,0.5,0.35} 
\definecolor{javapurple}{rgb}{0.5,0,0.35} 
\definecolor{javadocblue}{rgb}{0.25,0.35,0.75} 

\lstdefinestyle{basejava}{
  language=java,
  showstringspaces=false,
  basicstyle=\small\ttfamily,
  keywordstyle=\bfseries\color{javapurple},
  commentstyle=\itshape\blue,
  identifierstyle=\blue,
  frame=none,
  backgroundcolor=\color{white},
}

\lstdefinestyle{CustomJava}{
  belowcaptionskip=\baselineskip,
  breaklines=true,
  xleftmargin=\parindent,
  language=java,
  showstringspaces=false,
  basicstyle=\scriptsize\ttfamily,
  keywordstyle=\bfseries\color{javapurple},
  commentstyle=\itshape\blue,
  identifierstyle=\blue,
  belowskip=1pt,
  numbers=left,
  gobble=0
}

\lstdefinestyle{CustomJavaWoNumbers}{
  belowcaptionskip=0.5\baselineskip,
  breaklines=true,
  xleftmargin=\parindent,
  language=java,
  showstringspaces=false,
  basicstyle=\scriptsize\ttfamily,
  keywordstyle=\bfseries\color{javapurple},
  commentstyle=\itshape\blue,
  identifierstyle=\blue,
  belowskip=0.5pt,
  numbers=none,
  gobble=0
}

\lstdefinestyle{codit}{
  belowcaptionskip=\baselineskip,
  breaklines=true,
  language=java,
  showstringspaces=false,
  basicstyle=\scriptsize\ttfamily,
  keywordstyle=\bfseries\color{javapurple},
  commentstyle=\itshape\blue,
  identifierstyle=\blue,
}

\usepackage{listings}
\usepackage{xcolor}
\newcommand{\ic}[1]{\begin{small}\texttt{#1}\end{small}}
\definecolor{codegreen}{rgb}{0,0.6,0}
\definecolor{codegray}{rgb}{0.5,0.5,0.5}
\definecolor{codepurple}{rgb}{0.58,0,0.82}
\definecolor{backcolour}{RGB}{242,242,242}

\lstdefinestyle{mystyle}{
    language=Python, 
    backgroundcolor=\color{backcolour},   
    commentstyle=\bfseries\color{codegreen},
    keywordstyle=\bfseries\color{magenta},
    numberstyle=\footnotesize\color{codegray},
    stringstyle=\bfseries\color{codepurple},
    emph={l, sorted, x},
    emphstyle=\color{red},
    basicstyle=\ttfamily\footnotesize,
    breakatwhitespace=false,         
    breaklines=true,                 
    captionpos=b,                    
    keepspaces=true,                 
    numbers=left,                    
    numbersep=5pt,
    xleftmargin=10pt,
    showspaces=false,                
    showstringspaces=false,
    showtabs=false,                  
    tabsize=2,
    frame=lines,
    morekeywords={assert, True, False},
}

\lstdefinestyle{mystyle-simple}{
    language=Python, 
    backgroundcolor=\color{backcolour},   
    commentstyle=\bfseries\color{codegreen},
    keywordstyle=\bfseries\color{magenta},
    stringstyle=\bfseries\color{codepurple},
    emph={sort},
    emphstyle=\color{red},
    basicstyle=\ttfamily\footnotesize,
    breakatwhitespace=false,         
    breaklines=true,                 
    captionpos=b,                    
    keepspaces=true,
    showspaces=false,                
    showstringspaces=false,
    showtabs=false,                  
    tabsize=2,
    frame=lines,
    morekeywords={assert, True, False},
}

%% file: includes/pygments.tex
\newcommand\blue[1]{\textcolor[rgb]{0.00,0.00,1.00}{{#1}}}

\definecolor{blueish}{RGB}{250, 250, 255}
\definecolor{greenish}{RGB}{250, 255, 250}
\definecolor{redish}{RGB}{255, 200, 200}

\definecolor{highlight}{RGB}{175, 255, 100}
\definecolor{gray01}{gray}{.98}
\definecolor{gray05}{gray}{0.95}
\definecolor{gray08}{gray}{0.92}
\definecolor{gray10}{gray}{0.90}
\definecolor{gray12}{gray}{0.88}
\definecolor{gray15}{gray}{0.85}
\definecolor{gray18}{gray}{0.82}
\definecolor{gray20}{gray}{0.80}
\definecolor{gray25}{gray}{0.75}
\definecolor{gray30}{gray}{0.70}
\definecolor{gray35}{gray}{0.65}
\definecolor{gray40}{gray}{0.60}
\definecolor{gray45}{gray}{0.55}
\definecolor{gray50}{gray}{0.50}
\definecolor{gray55}{gray}{0.45}
\definecolor{gray60}{gray}{0.40}
\definecolor{gray65}{gray}{0.35}
\definecolor{gray70}{gray}{0.30}
\definecolor{gray75}{gray}{0.25}
\definecolor{gray80}{gray}{0.20}
\definecolor{gray85}{gray}{0.15}
\definecolor{gray90}{gray}{0.10}
\definecolor{gray95}{gray}{0.05}

%% file: includes/macros.tex
\xspace%

\newcommand{\Comment}[1]{}

\newtcbox{\inlinebox}[1][]{enhanced,
 box align=base,
 nobeforeafter,
 colback=blueish,
 size=small,
 left=0pt,
 right=0pt,
 boxsep=2pt,
 #1}

\newcommand{\RS}[2]{%
    \begin{result}
        \textbf{\hyperref[rq-#1]{Key Findings {#1}}~}{#2}%
    \end{result}
}

\renewcommand{\cref}[1]{\Cref{#1}}
\renewcommand{\ic}[1]
{\begin{small}\texttt{#1}\end{small}}
\newcommand{\gpt}{\ic{GPT-3.5-turbo}\xspace}
\newcommand{\gptfourt}{\ic{GPT-4-turbo}\xspace}
\newcommand{\gptfourl}{\ic{GPT-4-32k}\xspace}

\newcommand{\pass}{pass@}

\newcommand{\rom}[1]{\uppercase\expandafter{\romannumeral #1\relax}}

\newcounter{hypothesis}
\usepackage{listings}
\usepackage{color}

\definecolor{dkgreen}{rgb}{0,0.6,0}
\definecolor{gray}{rgb}{0.5,0.5,0.5}
\definecolor{mauve}{rgb}{0.58,0,0.82}

\newcommand{\rqa}{How does TiCoder impact the performance of python developers evaluating AI generated code, in terms of task correctness, time, and cognitive load?}

\newcommand{\rqc}{Does the TiCoder workflow improve the accuracy of generated code suggestions?}

\newcommand{\passk}[1]{\texttt{pass@#1}}
\newcommand{\TOOL}{\textsc{TiCoder}}
\newcommand{\tappProblem}{{\sc TiCoder}}

\newcommand{\txtDavinciThree}{\ic{text-davinci-003}}
\newcommand{\codex}{\ic{code-davinci-002}}
\newcommand{\codeGenSixB}{\ic{CodeGen-6B}}
\newcommand{\codeGenSevenB}{\ic{CodeGen2.5-7B}}

\newcommand{\prfx}{{\it prfx}}

\newcommand{\yesResponse}{{\sc Pass}}
\newcommand{\noResponse}{{\sc Fail}}
\newcommand{\dnResponse}{{\sc Undefined}}

\newcommand{\tappBool}{{\sc \tappProblem{}-PassFail}}
\newcommand{\tappOutput}{{\sc \tappProblem{}-Output}}

%% file: body/abstract.tex
Large language models (LLMs) have shown great potential in automating significant aspects of coding by producing natural code from informal natural language (NL) intent. 
However, given NL is informal, it does not lend easily to checking that the generated code correctly satisfies the user intent.

In this paper, we propose a novel interactive workflow \tappProblem{} for guided intent clarification (i.e., partial formalization) through tests to support the generation of more accurate code suggestions. Through a mixed methods user study with 15 programmers, we present an empirical evaluation of the effectiveness of the workflow to improve code generation accuracy. We find that participants using the proposed workflow are significantly more likely to correctly evaluate AI generated code, and report significantly less task-induced cognitive load. Furthermore, we test the potential of the workflow at scale with four different state-of-the-art LLMs on two python datasets, using an idealized proxy for a user feedback. We observe an average absolute improvement of 45.97\% in the \pass{1} code generation accuracy for both datasets and across all LLMs within 5 user interactions, in addition to the automatic generation of accompanying unit tests. 

%% file: body/introduction.tex
\section{Introduction}
\label{sec:intro}

\IEEEPARstart{L}{arge} Language Models (LLMs) have shown tremendous potential in generating natural-looking programs from informal intent expressed in natural language. 
There has been a surge in research around training LLMs over programming language artifacts in just the last couple of years~\cite{codex_2021, palm_2022, codegen_2022, incoder_2022, polycoder_2022}. 
Commercial offerings such as GitHub Copilot~\cite{copilot_2022} are widely available, and have been shown to generate a non-trivial fraction of code in real-world scenarios~\cite{copilot_maps_2022}. 

However, there are several challenges that arise when generating code from natural language specifications.\cite{liang2023understanding, xu2022ide}. For example, natural language prompts crafted by users may not always fully capture a their intent, as they may contain ambiguous language and lack of nuance. More importantly, it is not possible to automatically evaluate whether code generated from a natural language prompt is correct. Natural language is inherently ambiguous and enforcing the \emph{user intent} through some mechanical process (such as testing, static analysis or formal verification) is not immediately possible. 

Consider the following docstring, taken from MBPP~\cite{austin2021program}, a popular Python programming tasks benchmark: 
\begin{lstlisting}[style=mystyle]
def text_lowercase_underscore(text):
    """Write a function that returns true if the input string contains sequences of lowercase letters joined with an underscore and false otherwise""
\end{lstlisting}

While the intent may seem obvious at first, it is not immediately clear how to check the correctness of a potential solution. Querying an LLM such as \txtDavinciThree ~\cite{ouyang2022training} yields several plausibly correct code implementations that pass simple tests such as rejecting the empty string `` ", or accepting the string \ic{"aa\_bb"}.
However, it may also produce subtly buggy code solutions that accept strings such as  \ic{"aa\_bb\_cc"}, which is \emph{inconsistent} with the original user intent that expects the string to consist \emph{entirely} of two sequences of lowercase letters joined by an underscore (as defined by the accompanying hidden reference solution and the validation tests from MBPP).
In practice, this can often lead to users accepting code with subtle bugs while using LLMs~\cite{asare2022github, perry2022users}.
The apparent ambiguity in this particular docstring, and more importantly the informal nature of natural language, highlights the inability to immediately ascertain the correctness of the code generated by an LLM. Instead, it would be desirable to avoid surfacing such subtly incorrect codes by first clarifying, and partially formalizing, the user intent into a checkable specification.

This issue can be compounded when users are  presented with a list of candidate suggestions from LLMs, such as in the Copilot VSCode IDE suggestions pane, which can display up to 10 suggestions. Users often have to linearly scan the list of code suggestions, review them, and reject the incorrect ones until arriving at one that satisfies their intent. In such situations, subtle bugs may be overlooked, with significant downstream impacts. In fact, several recent works exploring developer-AI interaction have highlighted the need for mechanisms to facilitate verification of AI-generated code\cite{bird2022taking, xu2022ide}, such as those that allow users to use tests that disambiguate between the different code suggestions~\cite{jha-icse-10}.

However, prior research has shown that it can be difficult for users to manually provide a sufficient number of test cases to disambiguate suggestions upfront~\cite{lau2009programming}.

Inspired by findings around example generation and disambiguation techniques in Programming By Examples (PBE)~\cite{zhang2020interactive}, and recent emerging ability of LLMs to generate tests~\cite{lemieux2023codamosa,dinella2022toga,schafer2023adaptive} \textbf{in this paper, we propose \emph{leveraging  user-feedback} through LLM-generated tests to improve the trust and correctness of LLM-generated code}. 
Specifically, we propose the workflow of {\it test-driven interactive code generation} (\tappProblem{}) to  (a) clarify (i.e., partially formalize) user intent through generated tests, and (b) generate a {\it ranked} list of code that is consistent with such tests. 

Let us demonstrate a simple instantiation of this framework using the earlier example, where a user prompts an hypothetical LLM to generate code satisfying their natural language intent. Instead of directly displaying a list of plausible code suggestions, our framework \tappProblem{} would query the user with a question:

\begin{lstlisting}[style=mystyle-simple]
text_lowercase_underscore("aa_bb_cc") == True? 
\end{lstlisting}

Let us assume that the user answers 'no', since they expect only two sequences of lowercase letters, joined by one underscore, as mentioned earlier.
The workflow would likely query the user again with the following question:
\begin{lstlisting}[style=mystyle-simple]
text_lowercase_underscore("aa_bb") == True? 
\end{lstlisting}

If the user says 'yes', then the system would output the list of approved tests, as well as a set of semantically ranked code suggestions that are consistent with those tests. 
Once the user chooses a suggestion from such a list, it would generate code along with accompanying tests. 
\begin{lstlisting}[style=mystyle-simple]
def text_lowercase_underscore(text):
    return True if bool(re.search(r'^[a-z]_[a-z]+$', text)) else False
def test_text_lowercase_underscore():
    assert text_lowercase_underscore("aa_bb")== True
test_text_lowercase_underscore()
\end{lstlisting}

In the case of LLM-based code generation, the generated tests not only help make natural language intent more precise and prune incorrect suggestions generated by the LLM, but can also serves as debugging aid for remaining suggestions and regression tests for future code edits~\cite{copilot_maps_2022}.  

 While the proposed framework appears intuitive, \hl{it may not scale to more complex code generation tasks. For example, in cases where the user is unable to validate tests, e.g. for tests that require intricate testing frameworks, the a workflow may not be tenable.}  Furthermore, the utility of the interactive framework is contingent upon (a) the ability of LLMs to generate useful tests, and (b) the cost-benefit trade-off of the overhead of user interaction versus the benefit on pruning and ranking of code suggestions. 
 
 To this end, we seek to understand: \textbf{How does the proposed workflow impact the performance of developers evaluating AI generated code?} In addition, the proposed framework should scale, augmenting the code generation accuracy of several open and closed-source LLMs. Thus we also seek to answer: \textbf{Does proposed workflow augment the accuracy of code generation models?}

To answer these questions, we explore the effectiveness of our proposed framework through a (1) mixed-effects user study and (2) a large scale evaluation of the approach on two Python benchmarks for code generation. \textbf{This paper makes the following contributions:}
\begin{enumerate}[left=0pt,nosep]
\item We propose an interactive workflow, \tappProblem{}, for guiding user intent clarification through automatically-generated tests and improving code generation accuracy of LLMs. \tappProblem{}  leverages off-the-shelf LLMs for generating  code and tests, and provides a mechanism to check AI-generated code through user-approved tests.
\item We evaluate the effectiveness of \tappProblem{} by conducting a mixed-methods user study comparing two different variants of \tappProblem{} for generating and evaluating code suggestions, including a baseline condition representing existing developer-AI interaction workflows. We observe a significant reduction in cognitive effort reported by participants using either variant of \tappProblem{} over existing interaction mechanisms. 
\item We further evaluate the performance of the \tappProblem{} workflow at scale by simulating user feedback, using the reference code solution as an idealized proxy. \tappProblem{} is evaluated on on two Python datasets, MBPP and HumanEval, and a mixture of four open and closed sourced LLMs. We demonstrate that  \tappProblem{} contributes to improving the code generation accuracy of all LLMs considered. We observe an average absolute improvement of 45.73\%  in \passk{1} code generation accuracy within 5 user interactions across both benchmarks. \hl{In fact, we observe} \tappProblem{} \hl{can boost smaller model }\passk{1} \hl{ accuracy to levels comparable to much larger models, such as }\gptfourl, \hl{within just one user interaction. }

\end{enumerate}

%% file: body/related.tex
\section{Related Work}
\label{sec:related}

\subsubsection{Improving Code Generation Accuracy} Techniques for improving code generation accuracy is a rapidly growing field of work. Unlike the work proposed in this paper, these techniques do not consider user feedback, or guide users in clarifying their intent formally; we cover them briefly.

AlphaCode~\cite{alphacode_2022} and CodeT~\cite{codet_2022} both propose techniques to improve code generation accuracy by generating tests using LLMs, and then grouping code suggestions by the set of tests that they satisfy. CodeT~\cite{codet_2022} refines the approach by scoring tests and code suggestions simultaneously by prioritizing tests that satisfy many code suggestions and prioritizing codes that satisfy many tests. 
\hl{While there are similarities with CodeT in using LLM generated tests to rerank generated code that results in code generation accuracy on benchmarks, \tappProblem{} is complementary as one can apply \tappProblem{} after CodeT. But more importantly, we argue that a user cannot trust the generated code from CodeT any more than using LLM directly. This is because the user is still presented with a set of code suggestions. In contrast, with \tappProblem{}, we first formalize the user intent through tests allowing the user to constrain the code that the user will need to eventually sample from. \tappProblem{} also allows users to modify test output in one setting, which is not possible in the CodeT approach, where the tests are fixed throughout.}
As part of future work, we plan to explore if our approach may benefit from code and test ranking algorithms in CodeT. 

Similarly, work on program synthesis~\cite{gulwani_2017, solar_lezama_2009} generates code that satisfies a formal specification either expressed as a logical specification or input-output tests~\cite{gulwani2011automating}. Our work differs in that we use LLMs to generate code from informal specifications, i.e. natural language intent. However, it would be interesting for future work to leverage user-provided tests to improve the quality of code generation, as explored in recent works~\cite{jain2022jigsaw, prose_multimodal_2021}.
In this work, to evaluate our proposed approach at scale, we simulate user feedback using the code reference implementation as an idealized proxy, similar to prior works in oracle-guided inductive synthesis~\cite{jha-icse-10,jha-acta-17} and interactive program synthesis~\cite{le2017interactive, interactive_synth_2020} where an an oracle (reference implementation or users) is queried to identify the output for a given input. However, prior works in this area appeal to an automatic symbolic engine (such as a constraint solver~\cite{interactive_synth_2020} or automata construction~\cite{zhang2020interactive}) to generate distinguishing example inputs for a pair of programs, which is inconceivable for general purpose imperative programming languages such as Python.

\vspace{-2mm}
\subsubsection{Usability of AI Programming Assistants}

There exists several prior works exploring the usability of AI programming assistants. In this section, we focus on recent work that identifies challenges related to the expressing of intent and control over the generation suggestions of AI assistants.

Liang et al. \cite{liang2023understanding} identify that \emph{giving up on incorporating generated code}, and \emph{lack of ability to provide feedback}, are the most common usability issues encountered when using completion-based AI programming assistants. This often occurs because the code does not implement the desired functionality, participants do not know why certain code was generated and had trouble controlling the output to be aligned with their desired intent. 

McNutt et al. \cite{mcnutt2023design} enumerate a design space of interactions with code assistants, including how users should be able to disambiguate candidate programs or refine their initial specifications, echoing prior studies have indicated that disambiguation can be valuable in the context of assistants like GitHub Copilot \cite{barke2023grounded} and traditional program synthesis tools \cite{mayer2015user}.  Similarly, Xu et al. \cite{xu2022ide} explored challenges of IDE-based AI assistants, including how well specified the queries that users formulate are. They find that participants frequently have trouble expressing intent in their natural language queries to the assistant, and issues of under specification often relate to ambiguous instructions, such as omitting variable names. 

Mozannar et al. \cite{mozannar2022reading} identify 12 core activities associated with using GitHub Copilot and find that programmers often iterate on their prompts until they obtain the suggestion they desire, and spend a significant amount of time verifying code suggestions. In fact, recent work by Bird et al. \cite{bird2022taking} shows that as result of AI-powered tools, developer roles are shifting so that more time is spent time reviewing code than actually writing code. \hl{Several recent works}\cite{tianyi_2022, nguyen2022empirical,imai2022github} \hl{identify clear opportunities for improving the accuracy of LLM code generation techniques.} Our work builds upon observations of previous studies, and explores mechanisms to support code evaluation tasks.

%% file: body/method.tex
\section{Research Questions and Paper Organization}

We briefly introduce the research questions and discuss paper organization. In the following Section~\ref{subsec:workflow} we introduce our proposed approach: TiCoder. Then we answer two distinct research questions:

\begin{itemize}
    \item [\textbf{RQ1}] \textbf{\rqa} 
    To answer RQ1, we conduct a user study, where participants use AI assistants augmented with the TiCoder workflow. We evaluate the cost benefit tradeoff of the proposed approach on developer effort when evaluating AI generated code.
    \item [\textbf{RQ2}] \textbf{\rqc}
    To answer RQ2, we explore the code generation accuracy of LLMs augmented with the TiCoder workflow on two code generation benchmarks in python.
\end{itemize}

 The methodology, evaluation, and results of each research question are organized in the following sections: Sections~\ref{sec:userstudymethod} and \ref{sec:userstudyresults} describe the methodology and results for RQ1, and Section~\ref{sec:benchmarkresults} describes the methodology and results for RQ2. We separate methods and results of RQs into distinct sections for clarity.  We conclude with a Discussion (Sec.~\ref{sec:udiscussion}) of the implications of our work to the broader research community, and the Limitations of the presented experiments (Sec.~\ref{sec:threats}). 

\section{Proposed Approach: TiCoder}
\label{sec:method}

\begin{figure}[htbp]
    \centering
    \includegraphics[width=0.99\linewidth]{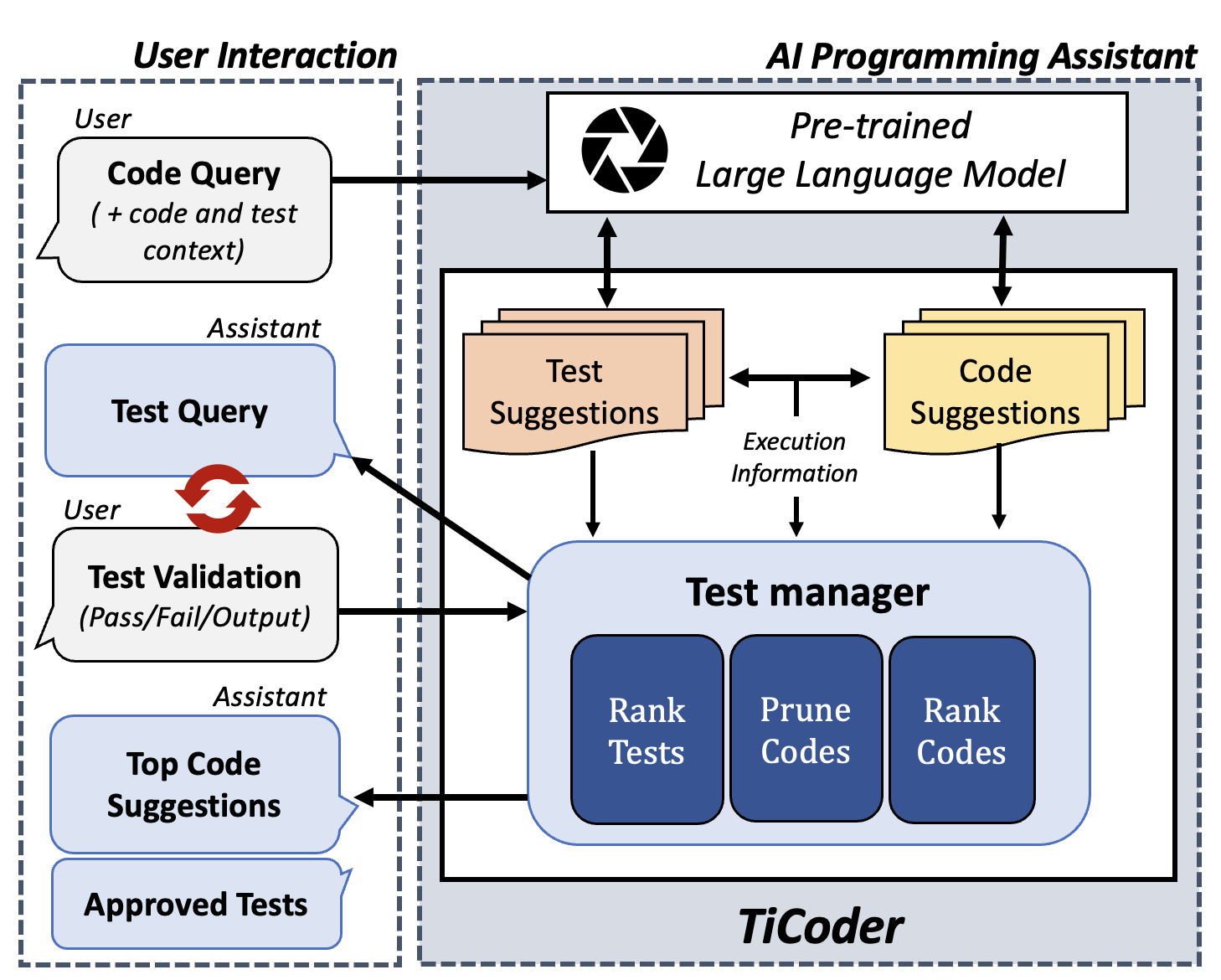}
    \caption{\tappProblem{} workflow.}
    \label{fig:tappy-approach}
\end{figure}

In this section, we outline a proposed workflow for leveraging test generation and user feedback to clarify (i.e., partially formalize)  user intent. We refer to this approach as \tappProblem{} ({\it Test-Driven Interactive Code Generation}), and define two variants of the workflow and surface this interaction to users in the following subsections.

\subsection{High-level Workflow}
\label{subsec:workflow}

Figure~\ref{fig:tappy-approach} describes the high-level workflow of {\it Test-Driven Interactive Code Generation} (\tappProblem{}).  

\setlist{nosep}

\begin{enumerate}[left=0pt,nosep]
    \item The user requests the AI programming assistant to generate a function, given optional code context including an existing prefix in a file, a natural language description, and the function header containing method name, parameters and returns.
    \item The AI programming assistant internally generates a set of candidate code and test suggestions by prompting an LLM.
    \item The set of generated tests are executed for each candidate code suggestion. The set of tests that pass or fail on each code suggestion are stored.
    \item Using execution information, the AI programming assistant {\it ranks} (according to some heuristics) the set of generated tests and then surfaces the top ranked test to the user as a query; asking the user if a test is consistent with the user's intent.
    \item The user responds either {\sc Pass}, \dnResponse{}, or {\sc Fail} signifying if the test is respectively: consistent,  precondition-violating\footnote{A test violates a precondition if the function is undefined on the test input. For example, the test \texttt{assert SquareRoot(-4) == -2}  undefined on negative numbers.}, or inconsistent with the user intent. 
    Optionally, in the case of {\sc Fail}, the user can provide the correct test output {\sc Output}.
    \item The AI programming assistant leverages the user response to prune, and rank the  set of code and test suggestions.  
    \item Interaction steps 4-6  can be repeated for multiple iterations, until a predefined termination criteria (e.g., fixed number of steps, absence of tests) has been satisfied. 
    \item Once the interaction terminates, the AI programming assistant outputs (a) a set of tests that the user has approved or specified, and (b) a ranked list of code suggestions that are consistent with the user responses.
\end{enumerate}

\textbf{We define two variants of the workflow:} \tappBool{} and \tappOutput{}. The first scenario represents the case where the user provides only a Boolean {\sc Pass,Fail}  response.
The second scenario, \tappOutput{}, extends the first scenario and represents the case where the user provides the expected output {\sc Output} in the case of a {\sc Fail} test.

We present both the scenarios as they enjoy complementary benefits. 
The \tappBool{} scenario is more lightweight, in terms of user feedback, as well as, generalizes well for richer tests beyond input-output examples. For example, tests for stateful APIs comprises of a test-prefix as input and the output oracle consists of a non-trivial predicate (e.g., checking functional correctness of a stack object using the predicate \texttt{s.pop() == a} on a stack object \texttt{s} and element \texttt{a})~\cite{dinella2022toga}. 
On the other hand, \tappOutput{} puts less burden on an LLM to create the correct output for a given test input; relying instead on the user. However, it may require the user to specify a possibly non-trivial test oracle when used beyond input-output examples.

\newcommand{\passat}[2]{{\tt pass@{#1}@{#2}}}
\newcommand{\passatbase}[1]{{\tt pass@{#1}}}
\newcommand{\acceptat}[1]{{\tt accept@{#1}}}
\newcommand{\positionCorrect}[1]{\tt PositionCorrect@{#1}}
\newcommand{\posCorrTests}{\tt PositiveTests}
\newcommand{\fractPosQ}[1]{\tt NumTests@{#1}}
\newcommand{\precision}[1]{\tt Precision@{#1}}
\newcommand{\numtries}{{\tt NumQueriesToAccept}}

\subsection{TiCoder Implementation}
\label{sec:ticoder-impl}
\begin{figure}[t]
    \centering
    \includegraphics[width=\linewidth]{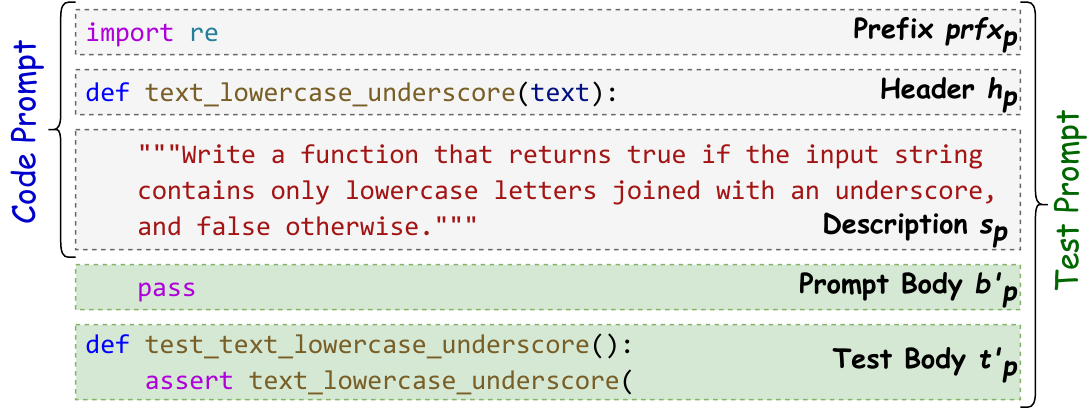}
    \caption{Example format, as well as \emph{code} and \emph{test prompts} for the running example.}
    \label{fig:mbpp-sample-prompts}
\end{figure}

In this section, we discuss one possible implementation of the \tappProblem{} workflow. 
Specifically, we outline the approach to generating code and test suggestions, ranking candidate tests to surface to the user, pruning and ranking code suggestions by user response. 
To simplify the presentation, we restrict ourselves to the case of single function synthesis, where the user input consists of a natural language comment $s_p$, the function header $h_p$, as well as any optional prefix $\prfx_p$ needed to generate the body of the function $p$. Figure~\ref{fig:mbpp-sample-prompts} shows an example for our running example. In addition, we also assume the presence of a set of {\it hidden} tests $T_p$ (input-output pairs for simplicity) to evaluate the correctness of the generated code, as well as a {\it hidden} reference (oracle) implementation of $p$, namely $b_p$. Our workflow does not have access to either $T_p$ or $b_p$.

\subsubsection{Generating Code and Tests}

We outline one possible choice for implementing the prompt generation for generating code and test suggestions for an example.

\newcommand{\ttt}[1]{{\tt #1}}

\begin{figure*}[htbp]
    \centering
    \includegraphics[width=0.97\linewidth]{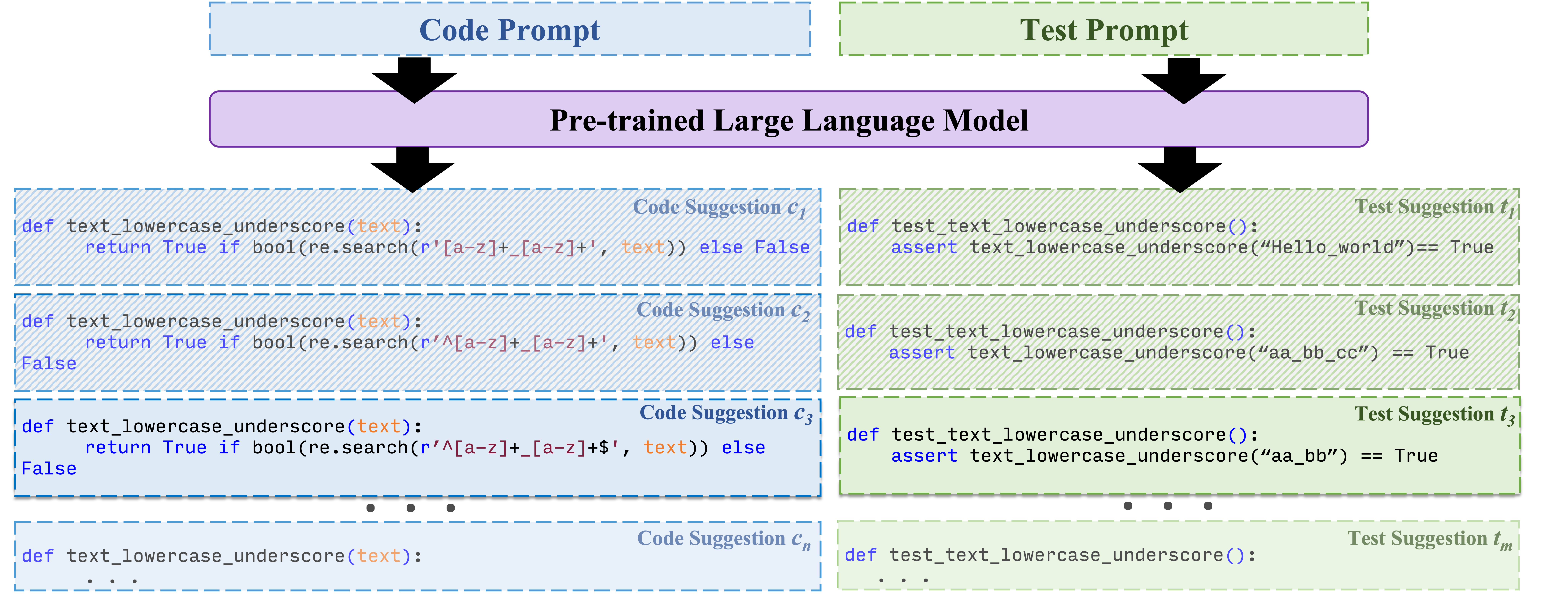}
    \caption{\emph{Code} and \emph{test suggestions} for the running example in \autoref{fig:mbpp-sample-prompts} generated from a LLM. Code suggestion $c_3$ and test suggestion $t_3$ are both \emph{correct}, while code suggestions $c_1$, $c_2$ and test suggestions $t_1$, $t_2$ are \emph{incorrect} (appear shaded), \emph{i.e.} they don't satisfy the \emph{problem prompts} in \autoref{fig:mbpp-sample-prompts}.}
    \label{fig:mbpp-sample-suggestions}
\end{figure*}
\autoref{fig:mbpp-sample-prompts} presents a possible \emph{code prompt} (in the gray boxes) that can be used to query an LLM to produce a set of \emph{code suggestions} for our running example. 
Querying a LLM with the code generation prompt will result in a set of \emph{code suggestions} similar to ones shown in \autoref{fig:mbpp-sample-suggestions}. 
Code suggestion $c_3$ is a valid solution to the problem, while $c_1$ is an incorrect code suggestion (since it allows the first substring to start with an uppercase letter) and $c_2$ is also incorrect (since it allows more than one sequence of lowercase letters joined with an underscore). Similarly, the green boxes in \autoref{fig:mbpp-sample-prompts} shows one possible \emph{test prompt} that augments the \emph{code prompt} with the statement \texttt{pass} as the method body (corresponding to a placeholder implementation in Python) along with the assertion to be completed within a test function. 
We use the generated test suggestions (\autoref{fig:mbpp-sample-suggestions}) to present the user with a set of tests.
Some of these are \emph{consistent} with the user intent ($t_3$); while  others are  \emph{inconsistent} with the user intent ($t_1$ and $t_2$). 

\subsubsection{Ranking test suggestions}
\label{sec:technique:testranking}

After obtaining the set of tests produced by an LLM, the user is presented with a sequence of tests. 
The user response to these proposed tests in both \tappProblem{} scenarios (\tappBool{}, \tappOutput{}) are used to prune and rank code suggestions. 
To minimize the  number of user interactions, it is desirable to prioritize tests that would result in the most number of incorrect code suggestions being pruned away~\cite{jha-icse-10,le2017interactive}. To achieve this, the set of tests are executed against the set of possible code suggestions generated by the LLM.

Then, using this execution information, we adopt a \ic{discriminative} test ranking policy that prioritizes tests that can discriminate best among the set of code suggestions generated by the LLM. 
If a test $t$ can discriminate between code suggestions well (i.e., splits the set of code suggestions into roughly equal halves), then it would prune away a substantial fraction of the code suggestions irrespective of the user response (either {\sc Pass} or {\sc Fail}). 

More precisely, let $U$ be the set of test suggestions and $G$ be the set of code suggestions that have not been pruned away after $k \geq 0$ user interactions.
For each test $t \in U$, we split the set of code suggestions $G$ into the sets $G^+_t$ and $G^-_t$ of code suggestions that pass and fail the assertion in $t$, respectively.
\hl{Note that we ignore codes that results in a crash on a test $t$ instead of failing with an assertion failure. We treat these as precondition violation.}
We then prioritize tests where the ratio of the sizes of these two set is closest to 1. 
In other words, we  rank the tests in decreasing order using the following scoring metric $s_{discr}$:

\begin{equation*}
\begin{split}
    s_{\textit{discr}}(t) = \begin{cases}
        0 & \text{if } \max(|G^+_t|, |G^-_t|) = 0, \\
        \frac{\min(|G^+_t|, |G^-_t|)}{\max(|G^+_t|, |G^-_t|)} & \text{otherwise}.
    \end{cases}
\end{split}
\end{equation*}



\vspace{2mm}

Note that the test ranking strategy is uniform for both the scenarios, although the test output will be possibly mutated by the user response in \tappOutput{}.


Consider the example in Figure~\ref{fig:mbpp-sample-suggestions}. 
Consider the two tests $t_1$ and $t_2$:
Two code suggestions \{$c_2$, $c_3$\} \ic{FAIL} on test suggestion $t_1$ while one suggestion \{$c_1$\}  \ic{PASS}, making $s_{\textit{discr}}(t_1) = \texttt{min}(1,2)/\texttt{max}(1,2) = 1/2$. Similarly, two code suggestions \{$c_1$, $c_2$\} \ic{PASS} on test suggestion $t_2$ while one suggestion \{$c_3$\}  \ic{FAIL} and $s_{\textit{discr}}(t_2) = 1/2$. 
All code suggestions in this example \ic{PASS} on test $t_3$ making $s_{\textit{discr}}(t_3) = 0$. 

\subsubsection{Pruning and ranking code suggestions}
\label{sec:rankcode}

\tappProblem{} returns a ranked list of code suggestions, whose behavior is consistent with all the user responses, and prunes the other code suggestions generated by the LLM, whose execution behavior on tests is contradictory to user expectation.
Let us first consider the case of code pruning.
Let $t \doteq (i,o)$ be a test in the form of an input-output example presented to the user.
If the user responds \yesResponse{}, then we prune any code $c \in G$ for which executing $c(i) \neq o$.
Similarly, if the user responds \noResponse{}, then we prune any code $c \in G$ for which executing $c(i) = o$.
In addition, for \tappOutput{} if the user provides the desired output $o'$ for the input $i$, then we can further prune any code suggestion $c$ for which $c(i) \neq o'$.
Note that we cannot soundly prune any code if the user responds with \dnResponse{}.

Finally, we define a simple  code ranking strategy that uses the tests in $U$ to determine a ranking on code suggestions in $G$ as follows:
Each generated code $c \in G$ is executed with each test $t \in U$ and gets assigned as a score \emph{the number of passing tests} $d_c$. The codes are then ranked based on the decreasing order of $d_c$. 

Following from the example in the previous section, represented in ~\autoref{fig:mbpp-sample-suggestions}, code suggestion $c_1$ passes on all tests \{$t_1$, $t_2$, $t_3$\}, code suggestion $c_2$ passes on \{$t_2$, $t_3$\} and code suggestion $c_3$ passes on \{$t_3$\}.
Our ranking would therefore rank $c_1$ highest initially in the absence of any feedback from the user.

%% file: body/userstudy.tex
\section{RQ1: User Study Methodology}
\label{sec:userstudymethod}
We aim to understand how the \tappProblem{} workflow may support software developers as they use AI-programming assistants to generate and evaluate code suggestions. We are seeking to answer the following research question:

\begin{itemize}
    \item [\textbf{RQ1}]\textbf{\rqa}
\end{itemize}

To answer our research question we conduct a controlled study with 15 participants consisting of 3 coding evaluation tasks. To complete each task, participants are asked to interact with one of the following AI assistants: Assistant 1 with no user intent refinement, Assistant 2 representing \tappBool{} workflow, or Assistant 3 representing \tappOutput{} workflow. Participants use each assistant to generate and evaluate a set of code suggestions.

We recruit participants using a mix of distribution lists and personal contacts. 3 of 18 participants were used as part of the pilot study to inform our design, and the remaining 15 are used in the final experiment. Table~\ref{tab:demo} contains participant demographic information. 8 participants are either professional software engineers or researchers at Microsoft, and the remaining 10 participants are PhD students from academia. The study was IRB approved with voluntary participation and paid \$15. All interviews were conducted over a video-conferencing platform and lasted approximately 45-minutes.

 Participants were asked to complete each code evaluation tasks with one of the three different AI code generation assistants. Each task had a time limit of 15 minutes. We use a within subject design, such that \emph{each participant uses all three assistants}, i.e. a different assistant for each task. Each AI assistant represents one treatment under study, which we describe in the next subsection.

\begin{table}[h!]
\resizebox{\linewidth}{!}{
\begin{tabular}{@{}lllll@{}}
\toprule
\multicolumn{1}{l}{ID} & \begin{tabular}[c]{@{}l@{}}Python\\ Experience\end{tabular} & \begin{tabular}[c]{@{}l@{}}Python\\ Frequency\end{tabular} & \begin{tabular}[c]{@{}l@{}}AI Programming \\ Assistant Use\end{tabular} & Occupation \\ \midrule
Pilot & \textgreater 5 years & Daily & Daily & Industry \\
Pilot  & \textgreater 5 years & Monthly & Monthly & Industry \\
Pilot  & \textgreater 5 years & Daily & Daily & Industry \\
P1 & \textgreater 5 years & Monthly & Daily & Industry \\
P2 & \textgreater 5 years & Weekly & Monthly & Industry \\
P3 & \textgreater 5 years & Rarely or never & Rarely or never & Industry \\
P4 & \textgreater 5 years & Daily & Daily & Industry \\
P5 & 3 - 5 years & Weekly & Weekly & Academia \\
P6 & \textgreater 5 years & Weekly & Weekly & Academia \\
P7 & \textgreater 5 years & Weekly & Monthly & Industry \\
P8 & \textgreater 5 years & Monthly & Rarely or never & Academia \\
P9 & \textgreater 5 years & Daily & Monthly & Academia \\
P10 & 1 - 2 years & Weekly & Daily & Academia \\
P11 & \textgreater 5 years & Daily & Daily & Academia \\
P12 & 3 - 5 years & Weekly & Rarely or never & Academia \\
P13 & \textgreater 5 years & Weekly & Rarely or never & Academia \\
P14 & 3 - 5 years & Rarely or never & Rarely or never & Academia \\
P15 & \textgreater 5 years & Daily & Daily & Industry \\ \bottomrule \\
\end{tabular}}
\caption{*Weekly denotes a few times a week, *Monthly denotes a few times a month.}
\label{tab:demo}
\end{table}

\vspace{-6mm}
\subsection{Treatments}
 The experiment includes one control condition and two distinct treatment conditions, implemented as different AI programming assistants. Each assistant differs in it's interaction mechanism with the developer and dictates the method in which to surface the final set of code suggestions shown to each user.  To ensure that the same set of codes is shown to all participants across treatments, we pre-select the prompt used to generate code suggestions. Second, to ensure we measure the impact of our dependent variables on only the process of evaluating AI generated code, we also restrict the ability to edit the AI generated code suggestions. The interaction framework of each Assistant is described below:

\subsubsection{\textbf{Control condition: AI Programming Assistant 1}} Assistant 1 represents the control condition for the experiment. Given the pre-selected prompt, Assistant 1 generates 5 code suggestions for the user, surfaced in a random order. Participants using Assistant 1 always see 5 unique code suggestions. We make this decision to reflect the current user experience scenario of several real-world AI code generation tools, such as GitHub Copilot's completion panel.  For example, the GitHub Copilot completion panel in VSCode shows the user up to 10 possible code suggestions at a time. Our decision is also informed by research pointing to the benefit of surfacing multiple code suggestions\cite{mcnutt2023design, mayer2015user,barke2023grounded, liang2023understanding}. We limit the maximum number of codes to 5 so as to allow the participant to complete each task within 15 minutes. 

\subsubsection{\textbf{Treatment condition: AI Programming Assistant 2}} 

Assistant 2 represents the {\sc TiCoder-PassFail}    (Sec.~\ref{subsec:workflow}) scenario, where a user provides instructions in the form of a prompt, and then the Assistant generates test cases that the user must validate. The user validates each test by  indicating if the test should pass or fail. Assistant 2 then uses the tests to prune any of the 5 code suggestions that differ in behaviour validated by the user. For example, if the user decides that the test should pass, only codes that pass the test are retained. These retained code suggestions are shown to the user, in random order.

\subsubsection{\textbf{Treatment condition: AI Programming Assistant 3}} 

Assistant 3 represents the {\sc TiCoder-Output} scenario (Sec.~\ref{subsec:workflow}). Instead of indicating whether a test should pass/fail (i.e. Assistant 2 interaction mechanism), users must provide the expected output of the test. Assistant 3 then uses the tests completed by the user to prune any of the 5 code suggestions that do not generate an output consistent with what the participant defined.

Both Assistants 2 and 3 use the tests to prune away generated codes that do not match the behaviour specified by the participant. We restrict the number of pruned codes in each task so that a participant using Assistant 2 or 3 will always see between 3-4 code suggestions if they correctly evaluate the tests shown to them. We make this decision to reflect the potential real-world scenario where \tappProblem{} is able to prune away at least 1 of the candidate code suggestions generated by an AI Assistant. However, a participant may see less than 3 code suggestions if they specify a contradictory or incorrect program behaviour through their answers to the tests. 

Participants interact with the assistants in an online survey platform, but they are able to copy and paste the generated code and tests into an IDE of their choice during the task. All code is formatted so as to not introduce external factors into the participants' time. The interactive nature of the AI Assistants in encoded into the survey logic, to mimic real-world execution and pruning of code suggestions based on a user's answers. Participants can maintain their view of the tests they validated in the survey throughout the task.

\begin{table*}[]
\resizebox{\linewidth}{!}{
\begin{tabular}{@{}llll@{}}
\toprule
Task & Task Name & Description & Treatments \\ \midrule
T1 & {\sc LowerUnderscore} & \begin{tabular}[c]{@{}l@{}}Write a function that returns true if the input string  consists of two sequences of lowercase letters joined \\ with a single underscore and false otherwise.\end{tabular} & A1, A2, A3 \\ \midrule
T2 & {\sc FirstMissing} & Write a function that finds the smallest missing number from a sorted list of integers, starting from 0. & A1, A2, A3 \\\midrule
T3 & {\sc MaxProduct} & \begin{tabular}[c]{@{}l@{}}Write a function to find the maximum product formed by multiplying numbers \\ of an increasing contiguous subsequence of that array. The sequence may include negative numbers.\end{tabular} & A1, A2, A3 \\\midrule
\end{tabular}}
\caption{Tasks included in the user study, derived from the MBPP dataset. A1 represents the control treatment, A2 represents the \tappBool{} treatment, and A3 represents the \tappOutput{} treatment.}
\label{tab:tasks}
\end{table*}

\subsection{Task Design}
We selected coding tasks that would satisfy the following criteria, for each of the three tasks: (1)  evaluating 5 AI-generated code suggestions could be completed in fifteen minutes, (2) there are syntactically valid but semantically incorrect code completions given by the LLM (\ic{GPT-3.5}) with a diversity of error types across tasks (3) they varied in problem domain and complexity, and (4) the LLM could generate reasonable tests that capture the diverse error types. 

\subsubsection{Identifying Task Candidates}

We select task candidates from the MBPP dataset~\cite{google_llm_2021}, a popular code generation benchmark, consisting of short Python functions designed to be solved by entry-level programmers. MBPP  provides a natural language instruction, a set of tests, and a ground truth code implementation for each problem. \hl{We cluster functions from MBPP based on problem domain, complexity as measured by cyclomatic complexity and size of the function in terms of lines of code. } From each cluster we identified a set of candidate functions for which we generated a code and test completions for using a LLM. We finally selected 3 problems for the code completion tasks that best satisfied the selection criteria. These problems represent three distinct styles: {\sc MaxProduct} is an algorithmic task involving dynamic programming, {\sc LowerUnderscore} involves using regex for string manipulation, and {\sc FirstMissing} involves a recursive binary search. The tasks are detailed in Table~\ref{tab:tasks}.


\begin{figure*}[htbp]
    \centering
    \includegraphics[width=1\linewidth]{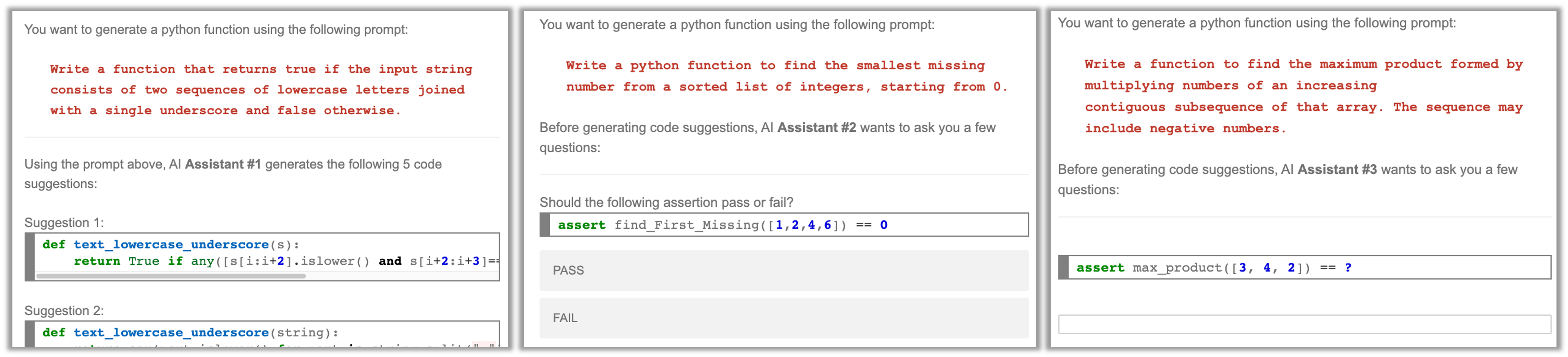}
    \caption{From left to right: Examples of different interaction sequences invoked by Assistant 1 on task T1 (directly display all code suggestions), Assistant 2 on T2 (validate the test output on a given input), and Assistant 3 on T3 (specify the output for a given input). }
    \label{fig:demo}
\end{figure*}
\subsubsection{Generating Code and Test Suggestions}

To generate the code and test candidates, we give the natural language instructions from the MBPP dataset as a prompt to the OpenAI \ic{GPT-3.5-turbo} chat completion endpoint with the default API parameters (temperature = 1.0).  We then sample a set of \hl{5 incorrect codes} using the tests from the MBPP dataset, to identify buggy programs. We also run the set of generated tests against the set of codes to make sure at least 1 and at most 2 code suggestions are caught by the test, to restrict the number of codes that would be pruned away. Rather than manually inject bugs into the ground truth program, we choose to sample the set of buggy codes from the LLM to reflect the nature of bugs users may encounter in AI generated code. 

The final set of tests and codes are fixed per task, regardless of the treatment used. For each task there are 5 suggestions: 4 buggy codes and 1 code that is extracted as the ground truth from the MBPP dataset. If the ground truth program extracted from MBPP does not handle certain pre-condition violations, we augment the code to match the task intent.  For Assistants 2 and 3, we show exactly 2 of the AI generated tests for each task. The final set of codes  are either directly shown to the user by Assistant 1, or first pruned based on the user's evaluation of the tests for Assistants 2 and 3.

\vspace{-4mm}

\subsection{Study Protocol}
At the start of the study, participants are given general instructions around how to interact with each AI assistant, the differences between them, and how to validate generated tests. Participants are also given time to set up their Python interpreter or environment before the start of the study. 
The survey interface used to interact with the AI assistants is shown in Figure~\ref{fig:demo}. Participants are able to view the coding task description, and depending on the treatment received, they can answer the AI Assistant's question, around the validation of a test case, directly in the survey. Once code suggestions are surfaced by the Assistant, participants are allowed to copy the code and run it for debugging, along with the set of provided tests, depending on the treatment. For each task, participants were asked to identify if the AI Assistant had returned a correct code suggestion, and if yes, which one. 

\hl{We employ a Latin Square Design to systematically vary the pairing of tasks and AI assistants. Each participant completes three tasks (T1, T2, and T3), each with a different AI assistant (Assistant 1, Assistant 2, Assistant 3). The order in which the participants use the AI assistants for each task is randomized to account for learning effects. This design ensures that each AI assistant is used an equal number of times across all tasks and positions, thus balancing potential order effects and providing a robust comparison of the AI assistants}

For each task, participants are encouraged to ask any questions around the task instructions. Our aim is to approximate the scenario where the user clearly understands what they want the AI Assistant to generate, such that they would query the AI Assistant with the same or similar prompt originally used to generate code and test suggestions. Although the real world usage of the workflow would differ, as developers often edit their prompts, we choose to fix the prompt to control for the generated code and tests across participants. Furthermore, we are not interested in the task of code generation, rather \emph{code validation}, i.e. not if the user can edit the prompt to get different suggestions, but rather how the \tappProblem{} workflow can help refine user provided natural language specification through tests, and how it may impact a developer's ability to validate code, and locate a correct suggestion out of a set of generated codes. While \tappProblem{} may help reduce the number of times a user must edit their original prompt, we save this exploration for future work.

\subsection{Measured Variables}
From the study recordings and user-submitted survey data, we collect a set of metrics on each task completed by the participants:

\emph{\textbf{Time.}} We measured time taken to complete each task from the recordings of each participant interview. Time for each task includes time taken to evaluate any tests. \hl{We measure time on task to determine if the \tappProblem{} workflow adds significant time overhead due to validation of tests, as compared to the control condition.}

\emph{\textbf{Correctness.}} The correctness is dependant on  1) the correct evaluation of the generated tests, relative to the oracle code implementation and 2) the correct selection of the code suggestion whose behaviour reflects the intent of the prompt. Each task has at most 1 correct answer: either one of the generated code suggestions is correct or none of suggestions are correct.

\emph{\textbf{Cognitive Load.}} \hl{The \tappProblem{} workflow aims to improve code generation accuracy and reduce the number of candidate code snippets that a user needs to review, ultimately reducing the cognitive effort of code evaluation. However, \tappProblem{} may also add additional cognitive effort, stemming from the effort required to validate tests. To test the impact of the workflow on cognitive effort,} after each task we measure participants’ self-reported cognitive load via their responses to five NASA TLX questions \cite{nasatlx}, a standardized approach to measuring self-reported cognitive load used widely across disciplines \cite{hart2006nasa}. We measure the following metrics using the standard 20 point scale: mental demand, effort, perceived success, pace, and stress.

\subsection{Evaluation of Measured Variables}

For each measured variable, time, correctness, and dimensions of cognitive load, we run a mixed-effects regression model. We use with either linear or logistic models depending on the data type. We use the treatment condition as the fix-effects independent variable, and participant ID and coding task as the random-effects variables.

We conduct an omnibus test using ANOVA to calculate the p-value of the treatment condition (the assistant used) against the measured metrics. To correct for multiple comparisons and conduct False Discovery Rate (FDR) correction\cite{benjamini1995controlling} for significant pairs of conditions. We only report Omnibus p-values for pairs of conditions for which the results are statistically significant, and the direction of significance. \hl{We chose mixed-effects models to account for individual variability (participants) and hierarchical data structures (task treatment pairs), using ANOVA for omnibus to test the significance of fixed effects (the assistant used), and conducted FDR to provide a comprehensive assessment of treatment effects while controlling for Type I errors.}

%% file: body/results.tex
\vspace{-2mm}
\section{RQ1: User Study Results}
\label{sec:userstudyresults}

 Our key quantitative results are summarized in Table~\ref{tab:rq1}.  The last column of Table~\ref{tab:rq1} provides Omnibus p-values for pairs of conditions for which the results are statistically significant.  
 \begin{table}[h]

\resizebox{1\linewidth}{!}{
\begin{tabular}{@{}lcccl@{}}
\toprule
\textbf{Metric} & \textbf{\begin{tabular}[c]{@{}c@{}}  Assistant 1\\  (mean)\end{tabular}} & \textbf{\begin{tabular}[c]{@{}c@{}}Assistant 2 \\ (mean)\end{tabular}} & \textbf{\begin{tabular}[c]{@{}c@{}}Assistant 3 \\ (mean)\end{tabular}} & \textbf{\begin{tabular}[c]{@{}c@{}}Pairwise \\ Significance\end{tabular}} \\ \midrule
\textbf{Correctness*}  (0,1) & 0.40 & 0.84 & 0.64 & \begin{tabular}[c]{@{}l@{}} $a1 \textless a2 (p = 0.001)$ \end{tabular} \\ \midrule
\textbf{Time} (seconds) & 327.7 & 284.15 & 253.88 & - \\ \midrule
\begin{tabular}[c]{@{}l@{}} \textbf{Cognitive} \\ \textbf{Load*} (0-100) \end{tabular} &  45.46 & 28.00 &  29.52 & \begin{tabular}[c]{@{}l@{}} $ a1 \textgreater a2 (p=0.007) $\\ $a1 \textgreater a3 (p=0.012)$ \end{tabular} \\ \midrule
\begin{tabular}[c]{@{}l@{}} \textbf{Mental} \\ \textbf{Demand*} (0-20) \end{tabular} & 12.5 & 7.50 & 7.6 & \begin{tabular}[c]{@{}l@{}}$a1 \textgreater a2 (p=0.001)$\\ $a1 \textgreater a3 (p=0.004)$\end{tabular} \\ \midrule
\textbf{Stress*} (0-20) & 8.26 & 3.84 & 6.35 & $a1 \textgreater a2 (p=0.02)$\\ \midrule
\textbf{Pace*} (0-20) & 8.13  & 5.38 & 4.70 &  $a1 \textgreater a3 (p=0.04)$ \\ \midrule
\textbf{Confidence} (0-20) & 13.5  & 15.92 &  15.88 & - \\ \midrule
\textbf{Effort*} (0-20) & 11.00  & 7.15 & 6.7  & \begin{tabular}[c]{@{}l@{}}$a1 \textgreater a2 (p=0.02)$\\ $a1 \textgreater a3 (p=0.014)$\end{tabular}  \\ \bottomrule \\
\end{tabular}}
\caption{Mixed-effects model analysis results for control (Assistant 1) and treatment (Assistant 2, 3) conditions.  \emph{(* denotes a significant observation. -- indicates no significance.)}}
\label{tab:rq1}
\end{table}

\label{sec:results}

\input{body/rqs/rq1}

%% file: body/rqs/rq1.tex
\subsection{\textbf{Impact on Task Correctness}} 

Using the mixed-effects regression model with the correctness of the task (coded as 0 or 1), as the dependent variable: The mean correctness was 0.40 for participants using Assistant 1, 0.84 Assistant 2, and 0.64 Assistant 3. Although the mean is higher in Assistant 2 and 3, the effect is significant for Assistant 2 only with (p=0.001). Looking at the set of mistakes made by participants, we notice several interesting observations. In general, participants using Assistant 1 are less likely to identify the correct code suggestion from the set of 5 suggestions.

For example, for Task 1, 3/4 participants that failed to identify the correct suggestion were using Assistant 1. Looking at their responses, all 3 participants identified different suggestions as correct. One participant, \textbf{P7} chose to not execute any of the code suggestions, while the other two participants \textbf{P3, P5} did write tests to evaluate the code suggestions, they were not able to find a test to characterize the bug. Similarly, for task 3, 2/4 participants that failed to identify the correct suggestion were using Assistant 1, and chose different candidate suggestions. Interestingly, both participants also only tested a subset of the codes, based on an initial guess of the correct suggestion.

Looking at the differences between Assistant 2 and 3, we notice that mistakes stem from both incorrect evaluations of the surfaced tests and incorrect evaluation of the code suggestions. For example, in Task 2 {\sc FirstMissing}, the first test case surfaced to participants by Assistant 2 is shown shown in Figure 2.b: \ic{assert find\_First\_Missing([1,2,4,6])==0}. All participants shown this test correctly answered that this test should pass. However, when the output \ic{== 0} is obfuscated on the same test by Assistant 3, 50\% of the participants indicated that the test should evaluate to 0, and the other 50\% indicated that it should (incorrectly) evaluate to 3. It is interesting to note that given a correct test by Assistant 2, participants are able to correctly evaluate it, however, if Assistant 2 had generated an incorrect test output ( == 3) it may not always be the case that participants are able to catch this bug.  

However, not all tests surfaced by Assistant 2 are correct. In Task 1, both tests surfaced by Assistant 2 had incorrect test outputs; testing edge case scenarios that should fail. For example \ic{text\_lowercase\_underscore("Hello\_world") == True}. For Assistant 2, all participants were able to correctly identify that the test should fail. For participants using Assistant 2, 4/5 indicated that it should evaluate to 'False' while one participant indicated that it should (incorrectly) evaluate to 'True'.

In Task 3, all participants using Assistant 2 and Assistant 3 were able to correctly evaluate the tests surfaced by the Assistants. However, 2 of the participants using Assistant 3 were not able to identify the correct code suggestion, whereas all participants using Assistant 2 were successful.

By construction, upon the erroneous evaluation of a test case by a user, the TiCoder workflow will prune all valid programs that pass on the test. Therefore, participants that incorrectly evaluate a test case will no longer see any valid AI-generated programs and cannot correctly complete the task, unless they specify that none of the code suggestions are correct. In the TiCoder workflow, noisy user response guarantees that generated code does not match the 'ground truth' user intended specifications. Therefore, in practice, the option to skip a test evaluation is imperative to the usability of the workflow, and to reduce uncertainty as the source of noisy input by the user. Though TiCoder may significantly support users in correctly evaluating code suggestions, the potential for noisy feedback is a critical risk to consider.

\RS{}{Participants using TiCoder Assistant 2 are significantly more likely to correctly evaluate AI generated code. Participants using Assistant 3 were, on average, more likely to correctly evaluate code suggestions compared to participants that were not using TiCoder. However, participants were also more prone to making mistakes while providing test outputs that dealt with edge cases.}

\subsection{\textbf{Impact on Task Time}} 

To test the effect of each Assistant on time, we used a mixed-effects regression model, with time as the dependent variable.
The mean time taken by participants using Assistant 1 is 327.7 seconds, 284.15 for Assistant 2, and 253.88 for Assistant 3. Although the means differ slightly across Assistants, on average participants using TiCoder take less time to complete the code evaluation tasks. However, this effect is not significant.

This indicates that the additional overhead of requesting participants to verify or provide the output for a test case does not add significantly to the time taken to complete the task. The time taken to evaluate code suggestions may be tempered by the number of code suggestions pruned, and the fact that Assistants 2 and 3 provide test cases to support the code evaluation process. One indicator of how long a participant takes to complete a task may be tied to their code evaluation strategy. We notice that, regardless of the treatment, participants that choose to execute and test every single code suggestion, take much longer than participants that scan the code suggestions and selectively execute and test candidate suggestions that 'look' correct to them. 

For example, (\textbf
{P2}) had relatively longer task times when using all 3 assistants, and chose to mentally execute every suggestion, identify the bug in each suggestion, and then proceeded to programmatically execute and test their hypothesis. Due to their thorough evaluation strategy, \textbf{P2} was correct on all tasks. However, we do not observe a correlation between time on task and correctness, both within, and across tasks (Pearson's Correlation Coefficient $r=0.016, p=0.911$).

\RS{}{ The time taken to validate test cases, introduced by TiCoder, does not introduce significant overhead to total task time. Participants using TiCoder take, on average, less time to complete the code evaluation tasks, however, this effect is not significant. }

\subsection{\textbf{Impact on Task Induced Cognitive Load}}
We analyze the self-reported cognitive load of participants across 5 dimensions, outlined by the NASA TLX: mental demand, effort, pace, stress, and confidence of the task correctness.  Cognitive load is reported as the cumulative sum across all 5 dimensions. Using a mixed-effects regression model with the cognitive load as the dependent variable, we observe that participants using Assistants 2 and 3 report significantly less cognitive load. Looking more closely at the different dimensions, for Assistant 2 participants report significantly less mental demand, stress, and effort required to complete the task.  For Assistant 3 participants report significantly less mental demand, effort, and better pace.

Overall, we posit that this effect might be observed due to the the reduced number of code suggestions that the user must evaluate, and that tests serve as concrete mechanisms for which to reason about the code; as well as provide a starting point for more extensive testing of the candidate functions, making it easier to get the task started. For example, when asking clarifying questions about the prompt used in a task, participants using Assistant 1 struggle to articulate their question before coming up with an illustrative test case. For Task 3 {\sc MaxProduct}, participants using Assistant 1 had difficulty conceptualizing 'increasing contiguous subsequence'. The interviewer made sure to answer any questions the participant had, but took care to not give concrete examples to not bias the participant. For example, \textbf{P19} first asked \emph{"so you're multiplying just two numbers, but it has to be next to each other?"}. When the interviewer clarified that it could be more than two numbers, given that the sequence is increasing, the participant articulated their question with an example \emph{"...so if I have, 1 2 4 1, it would be 1 by 2 by 4?"}. In contrast, participants that had similar questions, but were using Assistants 2 or 3, were able to more easily articulate their questions using test cases generated by the AI Assistant.

\RS{}{Participants using TiCoder, in both Assistant 2 and 3 settings, report significantly less task-induced cognitive load while evaluating AI generated code. This effect may be explained by the code pruning and test clarification mechanisms offered by TiCoder.}

%% file: body/benchmark.tex
\vspace{-2mm}
\section{RQ2: Benchmark Evaluation}
\label{sec:benchmarkresults}

Results from our user study, Section 5, indicate that TiCoder can significantly improve correctness of participants evaluating AI generated code, and that the workflow helps to reduce task-induced cognitive load. However, it is unclear if the proposed workflow is able to effectively generate tests that, once validated, can prune and rank a set of code suggestions with higher accuracy, on a large set of problems. To evaluate the potential utility of the TiCoder workflow at scale, we implement \tappBool{} and \tappOutput{}, and conduct an empirical evaluation on two state-of-the-art benchmarks for code generation in python. We aim to answer the following research question:

\begin{itemize}
    \item [\textbf{RQ2}] \textbf{\rqc}
\end{itemize}

\subsection{Datasets}
We use two Python programming datasets for our evaluation, including the {\it sanitized} version of the \emph{MBPP dataset} ~\cite{google_llm_2021}, dataset from Google, and the \emph{HumanEval dataset}, introduced in the Codex paper~\cite{codex_2021}, to answer the research questions.
\emph{MBPP} consists of 427 and \emph{HumanEval} of 164 examples in the format described in Sec~\ref{sec:ticoder-impl}, along with the hidden tests and reference implementations.
We modify the original \emph{HumanEval} dataset to remove any (non-hidden) input-output examples that are included in the docstring to avoid making the test generation task trivial.

\subsection{Evaluation metric}
For evaluating the {\it correctness of the generated code suggestions}, we use the popular metric  \passatbase{k} for evaluating the accuracy of code-generation by LLMs with respect to the hidden tests provided by each dataset~\cite{codex_2021}.
A code suggestion is correct if it passes all the hidden tests, and \passatbase{k} determines the mean {\it expected} value of \hl{an arbitrary sample of size $k$ to contain at least one correct solution.} 
To evaluate TiCoder, we define the metric \passat{k}{m} to denote the {\it ranked} \passatbase{k} for the code suggestions after $m \geq 1$ user queries.
Recall that TiCoder outputs a ranked list of code suggestions, so \passat{k}{m} measures if any of the top $k$ code suggestions is correct. 
\hl{Given that the list of code suggestions from TiCoder are ordered, our metric} \passat{k}{m} \hl{is not a statistical measure (unlike } \passatbase{k} \hl{which measures the statistical odds of any sample of size $k$ containing a correct code solution), but deterministically check if any one of the top $k$ ranked code suggestions is correct}.

\begin{table*}[ht!]
\resizebox{\linewidth}{!}{
\begin{tabular}{ccll|lllll|lllll} \toprule
 \multirow{3}{*}{Dataset} & \multirow{3}{*}{Model} & \multicolumn{2}{c|}{Baseline} & \multicolumn{5}{c|}{{\sc TiCoder-PassFail}} & \multicolumn{5}{c}{{\sc \hl{TiCoder-Output}}} \\  
 & \textbf{} & \multicolumn{2}{c|}{\textbf{pass@k}} & \multicolumn{5}{c|}{\textbf{pass@1@m}} & \multicolumn{5}{c}{\textbf{pass@1@m}} \\ \cline{3-14}
 &  & 1 & 100 & 1 & 2 & 3 & 4 & 5 & 1 & 2 & 3 & 4 & 5 \\ \cline{1-14}
\multirow{4}{*}{MBPP} & \txtDavinciThree & 49.16 & 86.88 & 68.04 & 75.26 & 77.33 & 77.88 & 78.08 & \hl{ 77.00 }& \hl{82.20} &  \hl{83.38} & \hl{83.59 } & \hl{83.75}\\
 & \codex & 48.25 & \blue{\textbf{89.75}} & 68.42 & 76.21 &  79.37 & 81.18 &  81.97 &  \hl{76.97} & \hl{85.51} &  \hl{87.10} & \hl{87.56 } & \hl{87.71}\\
 &  \codeGenSixB & 14.85 & 69.55 & 28.62 & 37.91 & 45.18 & 49.97 & 53.67 & \hl{39.64}  & \hl{54.01} &  \hl{62.34} & \hl{65.30}  & \hl{66.56}\\
   & \codeGenSevenB & 28.32 & 84.74 & 50.27 & 59.70 & 65.02 & 67.84 & 69.58 &  \hl{62.78} &  \hl{73.84} & \hl{78.10}  & \hl{79.52 } & \hl{80.42} \\ 
     & \hl{\gpt} & \hl{61.91 } & \hl{84.77}  &  \hl{75.12} & \hl{77.31}  & \hl{ 78.21}  & \hl{79.04}  &  \hl{79.03} & \hl{78.24}  & \hl{80.86 }& \hl{81.06}  & \hl{81.76}  & \hl{81.78} \\ 
  & \hl{\gptfourt} &  \blue{\textbf{\hl{69.80}}} &  \hl{86.88} & \hl{ 80.71} & \hl{81.90 } & \hl{82.38 } & \hl{83.20}  & \hl{83.12}  & \hl{83.91}  & \hl{84.63 }& \hl{84.97}  &\hl{85.65}   &  \hl{85.65}\\ 
 & \hl{\gptfourl} &  \hl{67.13} &  \hl{87.35} & \hl{ 81.56} &  \hl{82.62 }&\hl{ 82.97 } &  \hl{83.11 }&  \hl{83.70} & \hl{84.78}  & \hl{85.28 }&  \hl{85.31} & \hl{ 85.40} & \hl{85.79}\\
   \cline{1-14}
   
\multirow{4}{*}{HumanEval} & \txtDavinciThree & 44.13 & 87.80 &  60.70 &  67.54 &  71.41 & 72.18 & 72.81 & \hl{65.02}  & \hl{75.48} &   \hl{78.04} & \hl{79.34}  &\hl{80.18} \\
 & \codex & 30.49 & \blue{\textbf{91.49}} &  51.66 &  62.65 &   70.30 &  73.11 &  74.37 &  \hl{57.25} &  \hl{74.5} & \hl{81.99}  &  \hl{83.70} & \hl{84.50}\\
 & \codeGenSixB & 11.41 & 43.55 & 15.32 & 19.29 & 24.64 & 28.11 &  29.56 & \hl{26.34}  & \hl{32.81} &  \hl{39.08} & \hl{44.23}  & \hl{48.12}\\
  & \codeGenSevenB & 21.39 & 76.21 & 32.82 & 41.03 & 46.51 & 49.47 & 52.33 &  \hl{35.86} & \hl{51.60} & \hl{61.54}  & \hl{65.02}  &\hl{68.25} \\
  & \hl{\gpt} & \hl{59.45}  & \hl{89.02}  &  \hl{73.16} & \hl{76.48}  &  \hl{77.01} &  \hl{77.75} &   \hl{79.22} & \hl{73.44 }& \hl{76.60 }  & \hl{78.45 } &  \hl{78.45} &  \hl{79.51} \\ 
  & \hl{\gptfourt} & \blue{\textbf{\hl{62.62}}}  &   \hl{89.63} &  \hl{78.36} &  \hl{80.42} &  \hl{80.92} &  \hl{80.84} & \hl{81.34}  & \hl{82.22}  & \hl{83.17}  & \hl{83.42}  & \hl{83.62}   & \hl{83.84}  \\ 
 & \hl{\gptfourl} &  \hl{60.72} &  \hl{89.02} &  \hl{76.10} &  \hl{78.43} &  \hl{79.07} & \hl{79.48}  &   \hl{79.49} &    \hl{81.37}  &  \hl{82.46} &  \hl{82.49} & \hl{82.49} & \hl{82.54 }\\ 
  
  \bottomrule \\
\end{tabular}}
\caption{Model baseline and TiCoder results for two python datasets: MBPP and HumanEval. User interaction results are simulated for up to m=5 test evaluation interactions. We highlight, in blue, the highest accuracy in each column.  }
\label{tab:rq2}
\end{table*}

\subsection{Models and Baselines}
\tappProblem{} augments AI assistant workflows to improve the code generation accuracy of the underlying LLM. To assess \tappProblem{}'s benefits across various LLMs, we've chosen four state-of-the-art completion models, which include a mix of closed-source and open-source models. We provide a brief description of each model next.
\begin{itemize}

   \item  \emph{\textbf{OpenAI \codex{}, \txtDavinciThree{}}} \cite{codex_2021} is a closed source LLM specifically designed for code completion tasks. It is based on the GPT-3 architecture containing 175B parameters.\footnote{Access to this model was removed to the public by OpenAI in March 2023, but continues to be made free and available to researchers upon request.} \txtDavinciThree{} is also a closed source model of size 175B parameters, however, it is based on the GPT-3.5 architecture and InstructGPT~\cite{ouyang2022training} and can be used for a variety of natural language tasks. Compared to other non-chat based completion models, \txtDavinciThree{} demonstrates highly competitive performance on a number of tasks. 
   
 \item \emph{\textbf{ \hl{OpenAI \gpt{}, \gptfourt{}, \gptfourl}}} \hl{The OpenAI chat models are based on the pre-trained GPT-3 model, which is fine-tuned using Reinforcement Learning with Human Feedback (RLHF)}~\cite{ouyang2022training}. \hl{While these models are not explicitly fine-tuned for code generation, they have demonstrated strong capabilities on several related tasks}~\cite{olausson2023demystifying, fakhoury2023towards}. \hl{We use OpenAI APIs for the} \ic{gpt-3.5-turbo}, \ic{gpt-4-32k}, and \ic{gpt-4-turbo} endpoints.  

   \item  \emph{\textbf{Salesforce \codeGenSixB{}, \codeGenSevenB{}}} \cite{codegen_2022} is an open source LLM, with 6B parameters, trained specifically to translate natural language instructions to code.  \codeGenSevenB{} \cite{nijkamp2023codegen2} is an improvement on \codeGenSixB{} and is slightly larger, with 7B parameters. Currently, this model is the state-of-the-art for code generation compared to other models of similar parameter size. 
\end{itemize}

Our aim is to understand how TiCoder can help improve code generation accuracy across different LLMs, and not to identify the best performing model. Therefore,  we use default configurations for each model, and only alter temperature. 
We experimented with different temperature configurations to optimize performance and diversity of generated code and test suggestions. Intuitively, a temperature closer to 1 allows LLMs to provide a more diverse set of solutions, whereas a temperature closer to 0 forces LLMs to only generate fewer solutions with the highest confidence. 
\hl{Following}~\cite{codex_2021, li2023starcoder, roziere2023code} \hl{for all models we settle on a temperature of 0.8, as it maximizes the number of examples for which at least one correct code is produced within 100 suggestions for $k >1$ in $pass@k$}.
To account for the non-determinism of the LLM generations, for each dataset, we only query each model once to generate an initial set of 100 code and 50 test suggestions into a \emph{cache} of responses. We use the same cache across all experiments that involve the specific LLM.

\vspace{-2mm}
\subsection{Simulating User Response}
 
Our proposed workflow requires real-time user response to determine if a generated test is consistent with the user's intent. Therefore, in order to evaluate \TOOL{} {\it offline} with large-scale benchmark datasets, we define a proxy to {\it simulate} real-time user response. 
 
Similar to oracle-guided inductive synthesis~\cite{jha-icse-10,le2017interactive,jha-acta-17}, we use the reference implementation $b_p$ as an {\it oracle} to answer if a test $(i,o)$ is consistent with the user intent, and provide the expected output $b_p(i)$ when the test output $o$ does not match the user intent (for \tappOutput{}).
In other words, we assume that the intent of the user is precisely captured by the semantics of the (hidden) reference implementation.
\hl{Further, if a test input crashes the reference code, we treat the user response as} \dnResponse{} \hl{to model a precondition violation.}
However, this models an {\it idealized} user interaction because, in practice, users may sometimes be unable to answer a test query in a reasonable amount of time (say, when asked about the value of the 100th prime number). As observed in the results of the user study, unlike an idealized user, real participants may sometimes provide noisy input. For example, we observe that participants are more prone to making mistakes while providing test outputs that dealt with edge cases. Therefore, using the oracle as a proxy indicates that our empirical evaluation represents an {\it upper bound} on the improvement that TiCoder can have on the benchmarks with real users.

\input{body/rqs/rq3}

%% file: body/rqs/rq3.tex
\vspace{-3mm}
\subsection{Results}

To answer RQ2, we evaluate the performance of four different models, with and without TiCoder in the \tappBool{} and \tappOutput{} settings on MBPP and HumanEval datasets. Table~\ref{tab:rq2} contains all results for each model. 

The first column contains the baseline \passatbase{1} and \passatbase{100} for  each model on MBPP and HumanEval datasets. 
Note that \passatbase{100} denotes the fraction of examples for which an LLM generates at least one correct code suggestion within 100 suggestions.
The second and third columns contain the results for each model, augmented with TiCoder in \tappBool{} and \tappOutput{} settings respectively. We report the \passat{1}{m} metric, with m, the number of test-validation user interactions, ranging from 1 to 5. 
We report \passat{k}{m} only for the case of $k = 1$ as it is the strictest setting for assessing the impact of TiCoder. TiCoder improves the accuracy of \passat{k}{m} for higher values of $k$ as well, but we do not present them in the interest of space.

\hl{All three chat models,} \gpt{}, \gptfourt{} and \gptfourl \hl{demonstrate the highest accuracy for} \passatbase{1} \hl{ with comparable performance across datasets. As expected, }\txtDavinciThree{} and \codex{}, \hl{the two largest completion models, achieve the fourth and fifth baseline performance on both datasets. However, } \codex{} \hl{achieves the highest } \passatbase{100} \hl{across all models and both datasets}. 

Overall, we observe that both TiCoder in the \tappBool{} and \tappOutput{} settings significantly improve \passatbase{1} performance, across all models. As the number of test validation queries increase from $m=1$ to $m=5$ we also observe consistent improvement in \passatbase{1} performance. Although the improvement is most pronounced at $m=1$, compared to baseline.

For example, on MBPP, \tappBool{} improves \passatbase{1} baseline performance of \txtDavinciThree{} from 49.16\% to 68.04\%, an absolute improvement of 18.88\% within one user query. \tappOutput{} improves performance to 77.00\%, which is an absolute \passatbase{1} improvement of 27.84\% within one user query. This increases to 38.55\% with 5 queries. While smaller models achieve lower \pass{1} and \pass{100} accuracy, TiCoder still provides modest boosts in accuracy. For the worst performing model, \codeGenSixB{} on HumanEval, \tappBool{} provides an absolute \passatbase{1}  improvement of 3.91\% within one interaction, and \tappOutput{} provides an absolute improvement of \hl{14.93}\%.

\hl{In fact, we observe that} \tappProblem{} \hl{can boost code generation accuracy of smaller models to comparable performance of much larger SOTA LLMs. For example, after just one user interaction} \codex{} \hl{on MBPP achieves 68.42\% accuracy in the }\tappBool{} \hl{setting, which is in fact \emph{higher} than the }\passatbase{1} \hl{accuracy of all three SOTA chat models:} \gptfourl, \gptfourt, and \gpt.  

Finally, as expected \tappOutput{} consistently provides higher accuracy compared to \tappBool{}, since the former allows users to fix the incorrect test output.  \tappOutput{} achieves an average absolute improvement of 45.73\% in the code generation accuracy for both datasets and across all LLMs within 5 user interactions. However, it is worth noting that \tappBool{}, even with its lightweight feedback (that generalizes to richer tests or specifications), always stays within 9\% of the benefits of \tappOutput{}. 
This demonstrates the power of LLMs to generate test cases that satisfy user intent.

\RS{}{TiCoder significantly improves pass@1 performance for all studied LLMs on both benchmarks, with performance improvements increasing with every test validation interaction. \hl{TiCoder can boost small model performance within one user interaction, outperforming pass@1 accuracy of larger models like GPT-4-32k.} 
Additionally, the lightweight \tappBool{} scenario always stays within  9\% of the  performance of \tappOutput{} even in this idealized simulated user setting.}

%% file: body/threats.tex
\section{Discussion }
\label{sec:udiscussion}

\subsection{Tests as a developer-AI disambiguation mechanism}
Results of our user study indicate that using tests as an interactive mechanism to formalize user intent, and then prune and rank AI generated code suggestions, can meaningfully improve programmer performance. Results show that both Assistant 2 (\tappBool{}) and Assistant 3 (\tappOutput{}) are statistically distinguishable from Assistant 1 (the control condition without \tappProblem{}), where no interactive test case verification or code pruning is used. Participants using Assistant 2 are significantly more likely to correctly evaluate AI generated code suggestions, and report reduced task induced cognitive load, without negative impact on time to complete each task. However, compared to Assistant 2, participants made more mistakes when validating tests with Assistant 3.

Further research to explore a trade-off between the approaches, in practice, should be considered. However, we find that surfacing test cases in both forms might serve as a helpful mechanism for which to reason about generated code. While it is true that correctness likely depends on the evaluation strategies used by each participant, participants that were shown test cases by the AI Assistants performed significantly better.  

Recent work has shown that current developer-AI interaction workflows, as simulated by Assistant 1 in our study, have raised new issues in the way that developers write code. Results suggest a need for interaction mechanisms that can support disambiguation; a critical feature of the usability of AI-Assistants~\cite{mcnutt2023design,barke2023grounded}. We observed how \tappProblem{} can surface a ranked list of tests that delineate the space of possible code suggestions, providing a concrete mechanism for identifying potential ambiguities in natural language used to prompt LLMs. Furthermore, recent work has shown that developers spend significant amount of time verifying code suggestions~\cite{mozannar2022reading}\cite{bird2022taking}. While we do not observe a statistically significant impact on the time taken to evaluate AI-generated code suggestions when using \tappProblem{}, we do observe a significant reduction in the amount of cognitive effort required. We hypothesize that tests, which provide tangible artifacts for developers to reason about the code, can serve as a facilitating mechanism for constructing mental models of code functionality. Ultimately, this supports developers in the task of code evaluation. Future work should explore this in more detail.

\subsection{Improving LLM code generation capabilities with verified test cases}
Results of our benchmark evaluation indicate that, across all models studied, both implementations of \tappProblem{}, \tappBool{} and \tappOutput{}, can be used to augment the accuracy of an LLM through improved ranking and pruning. Specifically, we observe that using tests to better constrain the space of possible code suggestions can improve pass@k accuracy. This highlights the fact that current LLM capabilities may not be fully realized in practice: when prompted for multiple code suggestions, LLMs often are capable of generating a correct answer, but mechanisms to better rank the set of suggestions is needed. However, the performance boost provided by \tappProblem{} is contingent on a set of high quality tests used by \ic{discriminative} test ranking policy (Sec. ~\ref{sec:method}).

If the underlying LLM is unable to generate high quality tests, the ranking and pruning mechanism may not be as helpful. Future work should explore more sophisticated mechanisms for generating high quality tests that capture important specifications about the code. In this work, we explore the user study scenario where \tappProblem{} prunes away some, but not all of the AI generated code suggestions. 

It is worth noting that if the code suggestions generated by the underlying LLM predominantly exhibit consistent behavior, \tappProblem{} can still be valuable to a user by providing meaningful tests alongside the code suggestions. For example, in a scenario where all of the suggestions are correct with respect to the user's intent, \tappProblem{} may not prune away any of the code suggestions, but provides some guarantees about program behaviour to the user.

\subsection{The Value of Execution Based Pruning}

\begin{table}[h!]
\resizebox{\linewidth}{!}{%
\begin{tabular}{@{}lc|c|cc|cc@{}}
\toprule
\multirow{4}{*}{Dataset} & \multirow{4}{*}{Model} & \multirow{2}{*}{\sc{TiCoder-}} & \multicolumn{4}{c}{Tests in prompt} \bigstrut\\ \cline{4-7}
& & {\sc{PassFail}} & \multicolumn{2}{c|}{Single (pass@k)} & \multicolumn{2}{c}{All (pass@k)} \bigstrut\\\cline{3-7}
& & pass@1@1 & k = 1 & k = 5 &  k = 1 & k = 5 \bigstrut[t]\\\midrule
MBPP & \gptfourl & 81.56 & 78.26  & 87.81 & 80.88 & 91.58 \\ 
HumanEval & \gptfourl & 76.10 &   65.25 & 81.37   & 70.39 & 86.99 \\
\bottomrule
\end{tabular}
}
\caption{\hl{Results of prompting GPT-4 with validation tests in the prompt} }
\label{tab:testinprompt}
\end{table}

\hl{The }\tappProblem{} \hl{workflow helps users by automatically generating and ranking interesting test cases, instead of requiring the user to manually write them. However, users may have a set of test cases already in mind when querying a LLM}. \hl{We explore how such a scenario compares to the }\tappProblem{} \hl{workflow by adding the set of validation tests, i.e. the ground truth test set provided in each dataset, to the code generation prompt. We instruct }\gptfourl{} \hl{to generate a code suggestion that passes on the set of validation tests provided in the prompt. We then evaluate the $pass@k$ accuracy at $k=1$ and $k=5$ and compare to the }\tappProblem{} $pass@1@m$ accuracy from Table~\ref{tab:rq2}. Table~\ref{tab:testinprompt} \hl{contains the result of this experiment, showing both experiments where a single test is added in the prompt, as well as an experiment where all available validation tests are added in the prompt, the exact number of tests varies across both datasets.}

\hl{On both MBPP and HumanEval we observe a boost in $pass@k$ accuracy over the baseline prompt used Table}~\ref{tab:rq2}. \hl{For example, on MBPP} \gptfourl{} \hl{achieves a $pass@1$ of 67.13\% with the baseline prompt (no tests) and 78.26\% when including a single test in the prompt. This increases to a $pass@1$ of 80.88\% when all tests are added in the prompt. As expected, this demonstrates that adding tests in the prompt does help improve model performance. However, recalling the  $pass@1@1$ accuracy of the }\tappBool{}, \hl{within one user feedback loop, accuracy reaches 81.56\%, out performing the $pass@1$ accuracy where a single test is included in the prompt by 3.31\%. \tappProblem{} even performs slightly better (0.68\%) compared to the case when all tests are in the prompt. 

This indicates that even if the user supplies all test, there is no guarantee that the underlying LLM will fulfill the user intent and generate a code suggestion that passes on the tests. Conversely, with only one user interaction on a highly distinguishing test, obtained by ranking LLM generated tests, code generation accuracy greatly improves, \emph{matching the scenario where several tests are provided to the LLM without the added burden on the user to manually come up with test cases.} } 
 
\subsection{Considerations for AI-Generated Tests: Precondition Violations}
\hl{Tests generated by LLMs may contain pre-condition violating inputs that would cause the reference implementation to crash or fail, and would result in a} \dnResponse{} \hl{ response from the user, resulting in no pruning of code. 
 
}

\hl{As an illustrative case, consider from MBPP, the example of a reference implementation with (implicit) precondition that the argument \texttt{nums} array is non-empty, and throws a \texttt{divisionByZero} at the return statement for an empty \texttt{nums} array.}

\begin{lstlisting}[style=mystyle]
from array import array
def zero_count(nums):
   """Write a function to find the ratio of zeroes to non-zeroes in an array of integers."""
   n = len(nums)
   n1 = 0
   for x in nums:
      if x == 0:
         n1 += 1
      else:
         None
   return n1/(n-n1)
\end{lstlisting}

The following test is ranked highest by the discriminative ranking strategy:
\begin{lstlisting}[style=mystyle]
assert zero_count([]) == 0, "Empty List"
\end{lstlisting} 

For this test, out of the (deduplicated) 80 code suggestions from \codex, 8 suggestions pass the test, 11 suggestions fail the test and 61 suggestions crash on this test. 
The score for this test is 8/11 = 0.73, which is the highest among all tests.
\hl{However, this test does not lead to any code pruning as the user responds} \dnResponse{} in both the \tappBool{} and \tappOutput{} \hl{scenarios since the empty array causes the reference implementation above to fail.  Thus, our results in RQ2 account for tests with} \dnResponse{} \hl{responses, reflecting the possible real world impact of pre-condition violating tests on the accuracy boost provided by TiCoder.}

\section{Limitations and Threats}
\label{sec:threats}

\textbf{Generalizability of user study results.}
We evaluate \tappProblem{} under highly controlled experimental conditions, and the ability of developers to validate tests in more complex code generation scenarios may not scale. Our study explores two distinct test validation mechanisms surfaced in the \tappProblem{} workflow: \tappBool{} and \tappOutput{}. On the selected tasks, we observe that, in general, participants were able to successfully evaluate tests in both interaction scenarios. However, in practice specifying the output of generated tests may not always be a straightforward or simple task. In addition, we restrict participants abilities to edit the code prompt and code suggestions, to control for variables across participants. However, this is not a true reflection of real-world interaction behaviours. Future work should explore the impact of \tappProblem{} on developer productivity in real-world code settings, with broader audiences.  In addition, we only explored how \tappProblem{} impacts the correctness of code evaluation, i.e. how well users can disambiguate code suggestions. For example, a future experiment might examine how \tappProblem{} impacts online metrics such as code acceptance rates, or the total proportion of code contributed by the AI system, accommodating for solutions that provide partial correctness.

\textbf{Generalization of benchmark evaluation results.}
We also empirically evaluate \TOOL{} using two popular and state-of-the-art research Python benchmarks for code generation tasks: MBPP and HumanEval. While both benchmarks exercise common programming patterns, they may not be representative of real-world software development. Our findings may not generalize to a different set of programs across different languages and problem domains.

\textbf{Test execution overhead.} The proposed \tappProblem{} workflow incurs the cost of additional LLM inference, to generate candidate tests, as well as resource costs related to executing tests for generated code suggestions. Cost of execution may be non-trivial, and might not scale in scenarios where users wish to use an AI assistant to generate complex code. Nevertheless, the potential reliability guarantees and reduced effort for code verification represent a valuable trade-off when weighed against the costs of inference and execution in real-world scenarios. Future work should examine practical use of \tappProblem{} at scale.

%% file: body/conclusion.tex
\section{Conclusion}
\label{sec:conclusion}
 
In this work, we propose the workflow of test-driven interactive code generation using LLMs, and study it's effectiveness through a user study and empirical evaluation on code generation benchmarks. Our findings provide encouraging results around guiding user intent clarification for generating more correct programs.

In future work, we plan to extend and evaluate our implementation reflecting real-world scenarios including: more complex programs, an in-situ user study for various software development tasks, and an empirical evaluation on realistic benchmarks such as CoderEval~\cite{yu2023codereval} and NL2Fix~\cite{fakhoury2023towards}. 
  Finally, we plan to explore if \tappProblem{} can be extended to richer forms of formal specifications beyond tests, such as property based tests or pre- and post-conditions generated from user-defined prompts~\cite{endres2023formalizing}.